\documentclass[conference]{IEEEtran}
\IEEEoverridecommandlockouts
\usepackage{cite}
\usepackage{amsmath,amssymb,amsfonts}
\usepackage{algorithmic}
\usepackage{graphicx}
\usepackage{textcomp}
\usepackage{xcolor}
\usepackage[caption=false,font=footnotesize,subrefformat=parens]{subfig}
\def\BibTeX{{\rm B\kern-.05em{\sc i\kern-.025em b}\kern-.08em
    T\kern-.1667em\lower.7ex\hbox{E}\kern-.125emX}}
\begin{document}

\title{Versatile Compressive mmWave Hybrid Beamformer Codebook Design Framework
\\
}

\author{\IEEEauthorblockN{
Junmo Sung and Brian L. Evans}
\IEEEauthorblockA{\textit{Wireless Networking and Communications Group}, 
\textit{The University of Texas at Austin}
Austin, TX USA \\
junmo.sung@utexas.edu, bevans@ece.utexas.edu}
}

\maketitle

\begin{abstract}
Hybrid beamforming (HB) architectures are attractive for wireless communication systems with large antenna arrays because the analog beamforming stage can significantly reduce the number of RF transceivers and hence power consumption. In HB systems, channel estimation (CE) becomes challenging due to indirect access by the baseband processing to the communication channels and due to low SNR before beam alignment. Compressed sensing (CS) based algorithms have been adopted to address these challenges by leveraging the sparse nature of millimeter wave multi-input multi-output (mmWave MIMO) channels. In many CS algorithms for narrowband CE, the hybrid beamformers are randomly configured which does not always yield the low-coherence sensing matrices desirable for those CS algorithms whose recovery guarantees rely on coherence. In this paper, we propose a versatile deterministic HB codebook design framework for CS algorithms with coherence-based recovery guarantees to enhance CE accuracy. Simulation results show that the proposed design can obtain lower channel estimation error and higher spectral efficiency compared with random codebook for phase-shifter-, switch-, and lens-based HB architectures.   
\end{abstract}


\section{Introduction}

%
%
%
%

Millimeter wave multi-input multi-output (mmWave MIMO) communication is a promising technology for the next generation of cellular networks due to its potential enormous spectrum resource. MmWave bands have already been adopted in the 5G New Radio standard \cite{ts38104}, and multiple cellular service providers have made announcements on their mmWave deployment plans. 
Achieving large beamforming gains with large-scale antenna arrays is required to overcome the high propagation losses in these high frequency bands; however, in conventional all-digital beamforming systems, the many RF transceivers would bring about high power consumption due to power-demanding components therein. For reduction of power consumption and cost, hybrid analog and digital beamforming (HB) architectures have drawn attention as they reduce the number of RF chains while retaining a large number of antennas.
Channel estimation (CE) becomes challenging with HB architectures due to indirect access to entries of communication channel matrices and low signal-to-noise ratio (SNR) before transmit and receive beam alignment.


Because mmWave MIMO channels are sparse \cite{rapp2015tcom,rapp2013tap}, compressed sensing (CS) algorithms have been explored for sparse CE for HB \cite{Venu2017,gao2017twc,rf2018twc} as well as for analog beamforming \cite{rama2012ita} and digital beamforming \cite{schniter2014asilomar}. In those publications, transmit pilot symbols \cite{schniter2014asilomar,gao2017twc} or angles of phase shifters in analog beamformers \cite{rama2012ita, Venu2017,rf2018twc} are randomly chosen from finite predefined sets under hardware constraints. This random selection is expected to provide sensing matrices that are incoherent and satisfy the restricted isometry property (RIP) condition with high probability. A novel technique to generate and replace a dictionary matrix for CS formulation is proposed in \cite{xiao2019access}. Authors in \cite{lee2016tcom} instead proposed to design the pilot beam pattern by minimizing the total coherence of the equivalent dictionary for HB with phase shifters, which inspired the work in this paper. 
In addition to the CS-based approaches, codebook design for beam sweeping has also been explored, e.g., the low complexity broadened beam codebooks for all-digital MIMO \cite{raghavan2016beamforming} and the multi-resolution hierarchical codebook for HB MIMO \cite{Alkh2014stsp}.

The primary contribution of this paper is to generalize a specific class of open-loop mmWave channel estimation algorithms to work across the three most common hybrid beamforming architectures (phase shifter, switches, and RF lens).  The specific class of algorithms uses compressed sensing methods with recovery guarantees that rely on low coherence.  Our starting point is the design of a deterministic sensing matrix with low total coherence to obtain codebooks for analog and digital beamformers for hybrid beamforming using phase shifters \cite{lee2016tcom}.  We build on \cite{lee2016tcom} by replacing their approximation techniques to decompose the minimization problem into separate transmit and receive minimization problems with a derivation. We also build on \cite{lee2016tcom} by using a greedy approach to permute all columns of the equivalent dictionary for optimization, instead of using only a subset.  In simulations, we will use orthogonal matching pursuit (OMP) \cite{tropp2007jit} and basis pursuit denoising (BPDN) \cite{chen2001atomic} as representatives of CS algorithms with coherence-based recovery guarantees \cite{duarte2011tsp}.

\begin{figure*}[t!]
  \centering
  \includegraphics[width=12cm]{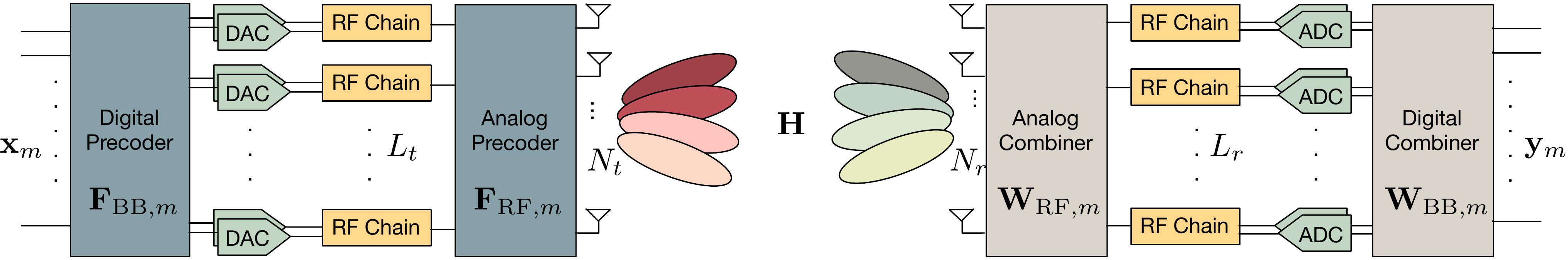}
  \caption{Block diagram of a general hybrid analog and digital beamforming architecture at both a transmitter and a receiver.}
  \label{fig:system_diagram}
  \vspace*{-0.15in}
\end{figure*}
\section{System and Channel Models} 
\label{sec:system_models}

We consider mmWave MIMO communication systems equipped with a general HB architecture as shown in Fig.~\ref{fig:system_diagram}. 
The analog beamforming stage can be implemented with various devices (e.g., phase shifters, switches and a DLA), and when phase shifters are taken into account, they have 
quantized angles with the $b_{\text{PS}}$-bit quantization resolution. A transmitter and a receiver are equipped with $N_t$ and $N_r$ antennas, and $L_t$ and $L_r$ RF chains, respectively, and it is assumed that $L_t \leq N_t$ and $L_r \leq N_r$. For frequency-flat channels, the discrete time received signal in the $m$-th frame (or the $m$-th time instant) can be written as
\begin{align}
  \mathbf{y}_m &= \sqrt{\rho} \mathbf{W}_m^{\mathsf{H}} \mathbf{H} \mathbf{F}_m \mathbf{x}_m + \mathbf{W}_m^{\mathsf{H}} \mathbf{n}_m \in \mathbb{C}^{L_r}, \nonumber
\end{align}
where $\rho$ is the transmit power in the training phase, $\mathbf{W}_m = \mathbf{W}_{\text{RF},m} \mathbf{W}_{\text{BB},m} \in \mathbb{C}^{N_r \times L_r}$ is the hybrid combiner, $\mathbf{H} \in \mathbb{C}^{N_r \times N_t}$ is the channel matrix, $\mathbf{F}_m = \mathbf{F}_{\text{RF},m} \mathbf{F}_{\text{BB},m} \in \mathbb{C}^{N_t \times L_t}$ is the hybrid precoder, $\mathbf{x}_m \in \mathbb{C}^{L_t}$ is the training symbols vector, and $\mathbf{n}_m \in \mathbb{C}^{N_r} \sim \mathcal{CN}(0, \sigma_n^2 \mathbf{I})$ is the additive noise vector. A hybrid combiner (precoder) in the $m$-th frame is composed of an RF combiner $\mathbf{W}_{\text{RF},m} \in \mathbb{C}^{N_r \times L_r}$ (an RF precoder $\mathbf{F}_{\text{RF},m} \in \mathbb{C}^{N_t \times L_t}$) and a baseband (BB) combiner $\mathbf{W}_{\text{BB},m} \in \mathbb{C}^{L_r \times L_r}$ (a BB precoder $\mathbf{F}_{\text{BB},m} \in \mathbb{C}^{L_t \times L_t}$). 
In order to keep the constant transmit power,  $\lVert \bar{\mathbf{x}}_m \rVert^2 = 1$ where $\bar{\mathbf{x}}_m \triangleq \mathbf{F}_m \mathbf{x}_m$. 
With $M_t$ and $M_r$ denoting the number of transmit and receive configurations, respectively, 
the received signal matrix $\mathbf{Y}$ that contains $M(=M_t M_r)$ frames can be written as
\begin{align}
  \label{eq:Y}
  \mathbf{Y} 
  &= \left[ [\mathbf{y}_1^{\mathsf{T}}, \ldots, \mathbf{y}_{M_r}^{\mathsf{T}} ]^{\mathsf{T}}, \ldots, [\mathbf{y}_{M - M_r + 1}^{\mathsf{T}} \ldots, \mathbf{y}_{M}^{\mathsf{T}}]^{\mathsf{T}}  \right] \nonumber \\
  &= \sqrt{\rho} \mathbf{W}^{\mathsf{H}} \mathbf{H} \bar{\mathbf{X}} + \mathbf{N},
\end{align} 
where $\mathbf{Y} \in \mathbb{C}^{L_r M_r \times M_t}$ is the received signal matrix, $\mathbf{W} \triangleq [ \mathbf{W}_1, \ldots, \mathbf{W}_{M_r} ] \in \mathbb{C}^{N_r \times L_r M_r}$ is the receive configuration matrix, $\bar{\mathbf{X}} \triangleq [ \bar{\mathbf{x}}_1, \ldots, \bar{\mathbf{x}}_{M_t} ] \in \mathbb{C}^{N_t \times M_t}$ is the transmit configuration matrix, and $\mathbf{N} \in \mathbb{C}^{L_r M_r \times M_t}$ is the noise matrix which is expressed as
$
  \mathbf{N} \triangleq \text{blkdiag}(\mathbf{W}_1, \ldots, \mathbf{W}_{M_r})^{\mathsf{H}} \times 
  \big[ [\mathbf{n}_1^{\mathsf{T}}, \ldots, \mathbf{n}_{M_r}^{\mathsf{T}}]^{\mathsf{T}}, \ldots, [ \mathbf{n}^{\mathsf{T}}_{M - M_r + 1}, \ldots, \mathbf{n}_{M}^{\mathsf{T}} ]^{\mathsf{T}} \big]  . \nonumber
$

Adopting a geometric channel model \cite{Alkh2014stsp,lee2016tcom} and linear antenna arrays, each scatterer contributes a channel path associated with its own azimuth angle of departure and arrival (AoD and AoA) denoted by $\theta_{tl}$ and $\theta_{rl}$, respectively. Therefore, the channel matrix can be expressed as
\begin{align}
  \label{eq:H_vector_form}
  \mathbf{H} = \sqrt{{N_t N_r}/{N_p}} \sum_{l=0}^{N_p - 1} \alpha_l \mathbf{a}_r(\theta_{rl}) \mathbf{a}_t^{\mathsf{H}}(\theta_{tl}),
\end{align}
where $N_p$ is the total number of paths (equivalent to the channel rank), $\alpha_l \sim \mathcal{CN}(0, \sigma_{\alpha}^2)$ is the complex channel gain of the $l$-th path, and $\mathbf{a}_t(\cdot) \in \mathbb{C}^{N_t}$ and $\mathbf{a}_r(\cdot) \in \mathbb{C}^{N_r}$ are the transmit and receive array response vectors, respectively, evaluated for the given angles. Both $\theta_{tl}$ and $\theta_{rl}$ are random variables that follow the uniform distribution $\mathcal{U}(0, 2\pi)$. Assuming the transmit and receive antennas are in the form of uniform linear array (ULA) with a half wavelength antenna spacing, the transmit array response vectors are given as
$
  \mathbf{a}_t(\theta) = \sqrt{{1}/{N_t}} \left[1, e^{-j \pi \cos(\theta)}, \ldots, e^{-j \pi (N_t - 1)\cos(\theta)} \right]^{\mathsf{T}},
$
and the receive array response is similarly defined. 
Defining the array response matrices 
$\mathbf{A}_t \triangleq [\mathbf{a}_t(\theta_{t0}), \mathbf{a}_t(\theta_{t1}), \ldots, \mathbf{a}_t(\theta_{t(N_p-1)})]$ 
and 
$\mathbf{A}_r \triangleq [\mathbf{a}_r(\theta_{r0}),\allowbreak \mathbf{a}_r(\theta_{r1}),\allowbreak \ldots,\allowbreak \mathbf{a}_r(\theta_{r(N_p-1)})]$, 
the channel matrix $\mathbf{H}$ in \eqref{eq:H_vector_form} can be expressed in matrix form as
\begin{align}
  \label{eq:H_matrix_form}
  \mathbf{H} = \mathbf{A}_r \mathbf{H}_d \mathbf{A}_t^{\mathsf{H}},
\end{align}
where $\mathbf{H}_d \in \mathbb{C}^{N_p \times N_p}$ is a square matrix with the scaled complex channel gains on the diagonal. 

\section{Sparse Formulation} 
\label{sec:sparse_formulation}
For application of CS algorithms to sparse CE, the received signal matrix in \eqref{eq:Y} can be rewritten in vector form using the matrix equality, $\text{vec}(\mathbf{A} \mathbf{B} \mathbf{C}) = (\mathbf{C}^{\mathsf{T}} \otimes \mathbf{A}) \text{vec}(\mathbf{B})$, as
\begin{align}
  \label{eq:y_1}
  \mathbf{y} = \sqrt{\rho} \left( \bar{\mathbf{X}}^{\mathsf{T}} \otimes \mathbf{W}^{\mathsf{H}} \right) \text{vec}(\mathbf{H}) + \mathbf{n} ,
\end{align}
where $\mathbf{y} \in \mathbb{C}^{M L_r}$ collects the $M$ received frames in vector form, and $\mathbf{n} \triangleq \text{vec}(\mathbf{N})$. 
In addition, angle grids need to be selected based on which CS algorithms search pairs of an AoD and an AoA.
We choose the angles to be uniformly distributed in the normalized discrete spatial angle domain, 
i.e., $[-1/2, 1/2)$,
such that $\vartheta_n \in \Theta = \{\vartheta_n | \vartheta_n = \frac{n}{G} - \frac{1}{2}, n = 0, \ldots, G-1 \}$ where $G \geq \text{max}(N_t, N_r)$ is the grid size. Therefore, the higher a $G$ value is, the finer the angle granularity becomes. The transmit array response vectors with spatial angle $\vartheta$ can be expressed as
  $
  \bar{\mathbf{a}}_t (\vartheta)= \sqrt{\frac{1}{N_t}} \left[ 1, e^{-j 2 \pi \vartheta}, \ldots, e^{-j 2 \pi (N_t - 1) \vartheta} \right]^{\mathsf{T}} \nonumber
  $
and the receiver one is also similarly defined. 
The angle grids are defined by the transmit and receive grid array response matrices $\bar{\mathbf{A}}_t \triangleq [\bar{\mathbf{a}}_t(\vartheta_0), \ldots, \bar{\mathbf{a}}_t(\vartheta_{G-1})] \in \mathbb{C}^{N_t \times G}$ and $\bar{\mathbf{A}}_r \triangleq [\bar{\mathbf{a}}_r(\vartheta_0), \ldots, \bar{\mathbf{a}}_r(\vartheta_{G-1})] \in \mathbb{C}^{N_r \times G}$, respectively. Note that $\bar{\mathbf{A}}_t \bar{\mathbf{A}}_t^{\mathsf{H}} = \frac{G}{N_t} \mathbf{I}_{N_t}$ and $\bar{\mathbf{A}}_r \bar{\mathbf{A}}_r^{\mathsf{H}} = \frac{G}{N_r} \mathbf{I}_{N_r}$. The channel matrix \eqref{eq:H_matrix_form} can be redefined with such grid array response matrices by
  $\mathbf{H} = \bar{\mathbf{A}}_r \bar{\mathbf{H}}_d \bar{\mathbf{A}}_t^{\mathsf{H}} \nonumber$
where $\bar{\mathbf{H}}_d \in \mathbb{C}^{G \times G}$ is the new channel gain matrix. Ignoring the grid quantization errors, $\bar{\mathbf{H}}_d$ is a sparse matrix with $N_p$ non-zero entries being complex channel gains corresponding to a combination of each transmit and receive array response vector in $\bar{\mathbf{A}}_t$ and $\bar{\mathbf{A}}_r$, respectively. Unlike $\mathbf{H}_d$, non-zero elements in $\bar{\mathbf{H}}_d$ do not have to be on the diagonal. As $\text{vec}(\mathbf{H}) = (\bar{\mathbf{A}}_t^{*} \otimes \bar{\mathbf{A}}_r) \text{vec}(\bar{\mathbf{H}}_d)$, the received signal vector in \eqref{eq:y_1} can be rewritten as
\begin{align}
  \mathbf{y} &= \sqrt{\rho} \left( \bar{\mathbf{X}}^{\mathsf{T}} \otimes \mathbf{W}^{\mathsf{H}} \right) \left( \bar{\mathbf{A}}_t^* \otimes \bar{\mathbf{A}}_r \right) \text{vec}\left( \bar{\mathbf{H}}_d \right) + \mathbf{n} \nonumber \\
  &= \sqrt{\rho} \left( \bar{\mathbf{X}}^{\mathsf{T}} \bar{\mathbf{A}}_t^* \otimes \mathbf{W}^{\mathsf{H}} \bar{\mathbf{A}}_r \right) \mathbf{h} + \mathbf{n} \nonumber \\
  &= \sqrt{\rho} \mathbf{\Phi} \mathbf{\Psi} \mathbf{h} + \mathbf{n}  = \sqrt{\rho} \bar{\mathbf{\Phi}} \mathbf{h} + \mathbf{n}, \nonumber 
\end{align}
where $\mathbf{h} \triangleq \text{vec}(\bar{\mathbf{H}}_d)$ is the $N_p$-sparse channel vector, $\mathbf{\Phi} \triangleq \bar{\mathbf{X}}^{\mathsf{T}} \otimes \mathbf{W}^{\mathsf{H}} \in \mathbb{C}^{L_r M_t M_r \times N_t N_r}$ is the sensing matrix, $\mathbf{\Psi} \triangleq \bar{\mathbf{A}}_t^* \otimes \bar{\mathbf{A}}_r \in \mathbb{C}^{N_t N_r \times G^2}$ is the sparsifying dictionary, and $\bar{\mathbf{\Phi}} \triangleq \mathbf{\Phi} \mathbf{\Psi}$ is the equivalent dictionary.

\section{Deterministic Sensing Matrix Design} 
\label{sec:deterministic_sensing_matrix_design}
Previous efforts have carefully designed a sensing matrix rather than using a random one in order to improve the performance of CS algorithms. Mutual coherence is one of the most popular metrics that are used for recovery guarantees of CS algorithms. 
Finding optimal sensing matrices using mutual coherence becomes intractable.
Alternatives to the mutual coherence, thus, are considered in \cite{ZM2011tsp,elad2007tsp,dc2009tip}. 

In \cite{lee2016tcom}, the authors considered minimizing the total coherence for sensing matrix design.  We build on \cite{lee2016tcom} by replacing their approximation techniques to decompose the minimization problem into separate transmit and receive minimization problems with a derivation.  We also build on \cite{lee2016tcom} by using greedy approach to consider all columns of the equivalent dictionary for optimization, instead of using only a subset.



 \subsection{Total Coherence} 
\label{sub:total_coherence}
We define the total coherence of the $\bar{\mathbf{\Phi}}$ as
\begin{align}
  \mu^t(\bar{\mathbf{\Phi}}) &= \sum_{m}^{G^2} \sum_{n, n \neq m}^{G^2} \left\lvert \bar{\mathbf{\Phi}}(m)^{\mathsf{H}} \bar{\mathbf{\Phi}}(n) \right\rvert^2 \nonumber \\
  &= \sum_{m}^{G^2} \left\{ \sum_{n}^{G^2} \left\lvert \bar{\mathbf{\Phi}}(m)^{\mathsf{H}} \bar{\mathbf{\Phi}}(n) \right\rvert^2 - \left\lvert \bar{\mathbf{\Phi}}(m)^{\mathsf{H}} \bar{\mathbf{\Phi}}(m) \right\rvert^2 \right\}. \nonumber 
\end{align}
As the total coherence is a sum of squared inner products of different columns in a matrix, $\mu^t(\bar{\mathbf{\Phi}})$ is the same as the squared Frobenius norm of $\bar{\mathbf{\Phi}}^{\mathsf{H}} \bar{\mathbf{\Phi}}$ without diagonal entries. Defining $\tilde{\mathbf{X}} \triangleq \bar{\mathbf{X}}^{\mathsf{T}} \bar{\mathbf{A}}_t^*$ and $\tilde{\mathbf{W}} \triangleq \mathbf{W}^{\mathsf{H}} \bar{\mathbf{A}}_r$, $\mu^t(\bar{\mathbf{\Phi}})$ can be expressed as
\begin{align}
  \label{eq:tc_Phibar}
  & \mu^t(\bar{\mathbf{\Phi}}) = \nonumber \\
  & \quad \sum_{m_1}^{G} \sum_{n_1}^{G} \left\lvert \tilde{\mathbf{X}}(m_1)^{\mathsf{H}} \tilde{\mathbf{X}}(n_1) \right\rvert^2 \sum_{m_2}^{G} \sum_{n_2}^{G} \left\lvert \tilde{\mathbf{W}}(m_2)^{\mathsf{H}} \tilde{\mathbf{W}}(n_2) \right\rvert^2 \nonumber \\
  & \quad - \sum_{m_1}^{G} \left\lvert \tilde{\mathbf{X}}(m_1)^{\mathsf{H}} \tilde{\mathbf{X}}(m_1) \right\rvert^2 \sum_{m_2}^{G} \left\lvert \tilde{\mathbf{W}}(m_2)^{\mathsf{H}} \tilde{\mathbf{W}}(m_2) \right\rvert^2.
\end{align}
As $\mu^t(\tilde{\mathbf{X}}) = \sum_{m}^{G} \sum_{n, n \neq m}^{G} \lvert \tilde{\mathbf{X}}(m)^{\mathsf{H}} \tilde{\mathbf{X}}(n) \rvert^2$ and $\mu^t(\tilde{\mathbf{W}}) = \sum_{m}^{G} \sum_{n, n \neq m}^{G} \lvert \tilde{\mathbf{W}}(m)^{\mathsf{H}} \tilde{\mathbf{W}}(n) \rvert^2$, \eqref{eq:tc_Phibar} can be rewritten as
\begin{align}
  \label{eq:total_coherence_Phi}
  \mu^t(\bar{\mathbf{\Phi}}) &= \mu^t(\tilde{\mathbf{X}}) \mu^t(\tilde{\mathbf{W}}) + \mu^t(\tilde{\mathbf{X}}) \nu(\tilde{\mathbf{W}}) + \mu^t(\tilde{\mathbf{W}}) \nu(\tilde{\mathbf{X}}),
\end{align}
where $\nu(\tilde{\mathbf{X}}) \triangleq \sum_{m}^{G} \lvert \tilde{\mathbf{X}}(m)^{\mathsf{H}} \tilde{\mathbf{X}}(m) \rvert^2$, and $\nu(\tilde{\mathbf{W}})$ is similarly defined. As all terms in \eqref{eq:total_coherence_Phi} are non-negative, minimizing $\mu^t(\bar{\mathbf{\Phi}})$ is split into four minimization problems for $\mu^t(\tilde{\mathbf{X}})$, $\mu^t(\tilde{\mathbf{W}})$, $\nu(\tilde{\mathbf{X}})$ and $\nu(\tilde{\mathbf{W}})$. 
Finding $\tilde{\mathbf{X}}$ in the transmitter and $\tilde{\mathbf{W}}$ in the receiver to minimize total coherence appear at first to be coupled, but due to \eqref{eq:total_coherence_Phi}, they become decoupled.  The following subsections will separately cover the minimization problems for the transmitter and the receiver.

\subsection{Transmitter Design} 
\label{sub:transmitter_design}
\subsubsection{Minimizing $\mu^t(\tilde{\mathbf{X}})$}
\label{subsub:minimizing_the_total_coherence_of_mu_X}
By definition of total coherence, $\min \mu^t(\tilde{\mathbf{X}})$ is equivalent to $\min \lVert \tilde{\mathbf{X}}^{\mathsf{H}} \tilde{\mathbf{X}} - \mathbf{I}_{G} \rVert_F^2$. 
The cost function can be expressed as
\begin{align}
  \label{eq:cost_function_Xtilde}
  \left\lVert \tilde{\mathbf{X}}^{\mathsf{H}} \tilde{\mathbf{X}} - \mathbf{I}_G \right\rVert_F^2 
  &= \left\lVert \tilde{\mathbf{X}} \tilde{\mathbf{X}}^{\mathsf{H}} - \mathbf{I}_{M_t} \right\rVert_F^2 + (G - M_t).
\end{align}
By definition of $\tilde{\mathbf{X}}$, $\tilde{\mathbf{X}} \tilde{\mathbf{X}}^{\mathsf{H}}$ can be rewritten as
$
  \bar{\mathbf{X}}^{\mathsf{T}} \bar{\mathbf{A}}_t^* \bar{\mathbf{A}}_t^{\mathsf{T}} \bar{\mathbf{X}}^* = \frac{G}{N_t} \bar{\mathbf{X}}^{\mathsf{T}} \bar{\mathbf{X}}^*
$.
Due to the transmit power constraint $\lVert \bar{\mathbf{x}}_m \rVert^2=1$, the diagonal elements of $\tilde{\mathbf{X}} \tilde{\mathbf{X}}^{\mathsf{H}}$ are fixed 
and cannot be minimized without reducing the transmit power. Thus off-diagonal elements are of our interest, and each $\tilde{\mathbf{X}} \tilde{\mathbf{X}}^{\mathsf{H}} (p,q)$ for $p \neq q$ can be written as
\begin{align}
  \label{eq:off_diagonal_element}
  \frac{G}{N_t} \mathbf{x}_p^{\mathsf{T}} \mathbf{F}_p^{\mathsf{T}} \mathbf{F}_q^* \mathbf{x}_q^* &= \frac{G}{N_t} \mathbf{x}_p^{\mathsf{T}} \mathbf{F}_{\text{BB},p}^{\mathsf{T}} \mathbf{F}_{\text{RF},p}^{\mathsf{T}} \mathbf{F}_{\text{RF},q}^* \mathbf{F}_{\text{BB},q}^* \mathbf{x}_q^*.  
\end{align}
Note that $p,q \leq M_t$ and that $M_t$ is desired to be large to reduce the lower bound shown in \eqref{eq:cost_function_Xtilde}. It leads to a larger matrix for which we would like to minimize the sum of squared off-diagonal elements. Here we have three sets of degrees of freedom to control the elements: the analog and digital precoders, and the training symbols. We define $\mathcal{F}_{\text{RF}}$ and $\mathcal{F}_{\text{BB}}$ as codebooks for possible analog and digital precoders, respectively.

Assuming $N_t$ is a multiple of $L_t$, we define the analog precoder codebook $\mathcal{F}_{\text{RF}}$ as
\begin{align}
  \mathcal{F}_{\text{RF}} &= \{ \mathbf{F}_{\text{RF},p} \in \mathbb{C}^{N_t \times L_t} : \forall p,q \in \{ 1, \ldots, N_t/L_t\}, \nonumber \\
  & \quad \mathbf{F}_{\text{RF},p}^{\mathsf{H}} \mathbf{F}_{\text{RF},p} = \mathbf{I}_{L_t}, \mathbf{F}_{\text{RF},p}^{\mathsf{H}} \mathbf{F}_{\text{RF},q} = \mathbf{0}_{L_t}, p \neq q \}. \nonumber
\end{align}
The first condition of the codebook ($\mathbf{F}_{\text{RF},p}^{\mathsf{H}} \mathbf{F}_{\text{RF},p} = \mathbf{I}_{L_t}$) is to facilitate further optimization. The second ($\mathbf{F}_{\text{RF},p}^{\mathsf{H}} \mathbf{F}_{\text{RF},q} = \mathbf{0}_{L_t}$) is to make off-diagonal elements zeros when different analog precoders are used. One obvious example of such $\mathcal{F}_{\text{RF}}$ is a set of submatrices of which each contains $L_t$ columns of the $N_t \times N_t$ discrete Fourier transform (DFT) matrix. 

Even if both $\mathbf{F}_{\text{RF},p}$ and $\mathbf{F}_{\text{RF},q}$ use the same analog precoder which does not necessarily make $\tilde{\mathbf{X}} \tilde{\mathbf{X}}^{\mathsf{H}} (p,q)$ zero, the digital precoder and the training symbol vector can be exploited to minimize such an element. Assuming $\mathbf{F}_{\text{RF},p} = \mathbf{F}_{\text{RF},q} \in \mathcal{F}_{\text{RF}} $ and $\mathbf{F}_{\text{BB},p} = \mathbf{U} \mathbf{\Sigma}_p \mathbf{V}^{\mathsf{H}}$ by the singular value decomposition (SVD) for all $p$, \eqref{eq:off_diagonal_element} can be written as
\begin{align}
  \label{eq:bb_precoder_config}
  (G/N_t) \mathbf{x}_p^{\mathsf{T}} \mathbf{F}_p^{\mathsf{T}} \mathbf{F}_q^* \mathbf{x}_q^* &= (G/N_t) \tilde{\mathbf{x}}_p^{\mathsf{T}} \mathbf{\Sigma}_p^{\mathsf{T}} \mathbf{\Sigma}_q^* \tilde{\mathbf{x}}_q^*, 
\end{align}
where $\tilde{\mathbf{x}}_p \triangleq \mathbf{V}^{\mathsf{H}} \mathbf{x}_p$ is the transformed training symbol vector, $\mathbf{\Sigma}_p$ is a diagonal matrix with diagonal entries being non-negative real numbers, and $\mathbf{U}$ and $\mathbf{V}$ are $L_t \times L_t$ arbitrary unitary matrices. To make \eqref{eq:bb_precoder_config} zero, we can choose one of the vectors for the diagonal of $\mathbf{\Sigma}_p$ such that they do not overlap one another. Since the elements in $\mathbf{\Sigma}_p$ are restricted to be non-negative real numbers, a straightforward option is to choose $L_t$-dimensional standard bases so that $\mathbf{\Sigma}_p^{\mathsf{T}} \mathbf{\Sigma}_q^* = \mathbf{0}_{L_t}$. The standard bases can be scaled; however, the digital precoders should meet the constraint $\lVert \mathbf{\Sigma}_p \tilde{\mathbf{x}}_p \rVert^2 = 1$. The constraint comes from the transmit power constraint $\lVert \mathbf{F}_{\text{RF},p} \mathbf{F}_{\text{BB},p} \mathbf{x}_p \rVert^2 = 1$ and the analog precoder codebook condition $\mathbf{F}_{\text{RF},p}^{\mathsf{H}} \mathbf{F}_{\text{RF},p} = \mathbf{I}_{L_t}$. 
It leads $\tilde{\mathbf{x}}_p$ to be the same standard basis that is used for $\mathbf{\Sigma}_p$ with a reciprocal of the scaler if scaled. It shows that the digital precoder and the training symbols have one-to-one correspondence, that the degrees of freedom we actually have is two, and that $\lvert \mathcal{F}_{\text{BB}} \rvert = L_t$. Thus, without loss of generality, we use the standard bases without scaling. Consequently the digital precoder codebook $\mathcal{F}_{\text{BB}}$ can be expressed as
\begin{align}
  \mathcal{F}_{\text{BB}} &= \big\{ \mathbf{F}_{\text{BB},p} \in \mathbb{C}^{L_t \times L_t} : \forall p \in \{ 1,\ldots,L_t \}, \nonumber \\
  & \quad \mathbf{F}_{\text{BB},p} = \mathbf{U} \mathbf{\Sigma}_p \mathbf{V}^{\mathsf{H}}, \mathbf{\Sigma}_p = \text{diag}(\mathbf{e}_p) \big\}, \nonumber
\end{align}
where $\mathbf{e}_p$ denotes the $L_t$-dimensional standard basis with one in the $p$-th entry. For the given $\mathbf{F}_{\text{BB},p}$, the corresponding training symbol vector is given as $\mathbf{x}_{p}=\mathbf{V} \mathbf{e}_p $.

\subsubsection{Minimizing $\nu(\tilde{\mathbf{X}})$}
\label{subsub:minimizing_nu_X}
We now focus on minimizing $\nu(\tilde{\mathbf{X}})$. As all terms in summation in $\nu(\tilde{\mathbf{X}})$ are non-negative, the problem $\min \nu(\tilde{\mathbf{X}})$ is equivalent to
$
  \min \sum_{m}^{G} \lvert \tilde{\mathbf{X}}(m)^{\mathsf{H}} \tilde{\mathbf{X}}(m) \rvert, \nonumber
$
and the cost function can be rewritten as
$
  \text{Tr}(\tilde{\mathbf{X}} \tilde{\mathbf{X}}^{\mathsf{H}}). \nonumber
$
It is a sum of the diagonal elements in $\tilde{\mathbf{X}} \tilde{\mathbf{X}}^{\mathsf{H}}$ which, as shown previously, are constrained by the transmit power. It means that $\nu(\tilde{\mathbf{X}})$ cannot be minimized without reducing the transmit power. 

\subsection{Receiver Design} 
\label{sub:receiver_design}
\subsubsection{Minimizing $\mu^t(\tilde{\mathbf{W}})$} 
\label{subsub:minimizing_the_total_coherence_of_mu_W}
As with the minimization of $\mu^t(\tilde{\mathbf{F}})$, $\min \mu^t(\tilde{\mathbf{W}})$ is equivalent to $\min \lVert \tilde{\mathbf{W}}^{\mathsf{H}} \tilde{\mathbf{W}} - \mathbf{I}_G \rVert_F^2$. As with \eqref{eq:cost_function_Xtilde}, we have the cost function given by
$
  \lVert \tilde{\mathbf{W}} \tilde{\mathbf{W}}^{\mathsf{H}} - \mathbf{I}_{M_r L_r} \rVert_F^2 + (G \allowbreak - M_r L_r),
$
and $\tilde{\mathbf{W}} \tilde{\mathbf{W}}^{\mathsf{H}}$ can be expressed as
\begin{align}
  \label{eq:WWH}
  \tilde{\mathbf{W}} \tilde{\mathbf{W}}^{\mathsf{H}} &= \frac{G}{N_r}
  \begin{bmatrix}
    \mathbf{W}^{\mathsf{H}}_{1} \mathbf{W}_1 & \cdots & \mathbf{W}^{\mathsf{H}}_{1} \mathbf{W}_{M_r} \\
    \vdots & \ddots & \vdots \\
    \mathbf{W}^{\mathsf{H}}_{M_r} \mathbf{W}_1 & \cdots & \mathbf{W}^{\mathsf{H}}_{M_r} \mathbf{W}_{M_r}
  \end{bmatrix}. 
\end{align}
We assume that $N_r$ is a multiple of $L_r$. Denoting $\mathcal{W}_{\text{RF}}$ and $\mathcal{W}_{\text{BB}}$ by codebooks for possible analog and digital combiners, respectively,
we define the analog combiner codebook as 
\begin{align}
  \mathcal{W}_{\text{RF}} &= \big\{ \mathbf{W}_{\text{RF},p} \in \mathbb{C}^{N_r \times L_r}: \forall p, q \in \{ 1, \ldots, {N_r}/{L_r} \} \nonumber \\
  & \quad \mathbf{W}_{\text{RF},p}^{\mathsf{H}}\mathbf{W}_{\text{RF},p} = \mathbf{I}_{L_r}, \mathbf{W}_{\text{RF},p}^{\mathsf{H}}\mathbf{W}_{\text{RF},q} = \mathbf{0}_{L_r}, p \neq q \big\}, \nonumber
\end{align}
for the same reason as with analog precoders. Off-diagonal blocks in \eqref{eq:WWH} become zero matrices due to the second condition. Focusing on the diagonal blocks in \eqref{eq:WWH}, it reduces to the following minimization problem:
$
  \min \left\lVert G/N_r \mathbf{W}_{\text{BB},p}^{\mathsf{H}} \mathbf{W}_{\text{BB},p} - \mathbf{I}_{L_r} \right\rVert_F^2 .
$
With $\mathbf{W}_{\text{BB},p} = \tilde{\mathbf{U}} \tilde{\mathbf{\Sigma}}_p \tilde{\mathbf{V}}^{\mathsf{H}}$ for all $p$ by the SVD, the cost function can be expressed as
\begin{align}
  \left\lVert ({G}/{N_r}) \tilde{\mathbf{V}} \tilde{\mathbf{\Sigma}}_p^{\mathsf{H}} \tilde{\mathbf{\Sigma}}_p \tilde{\mathbf{V}}^{\mathsf{H}} - \mathbf{I}_{L_r} \right\rVert_F^2 &= \left\lVert ({G}/{N_r}) \tilde{\mathbf{\Sigma}}_p^{\mathsf{H}} \tilde{\mathbf{\Sigma}}_p - \mathbf{I}_{L_r} \right\rVert_F^2 , \nonumber
\end{align}
where $\tilde{\mathbf{\Sigma}}_p$ is a $L_r \times L_r$ diagonal matrix with diagonal entries being non-negative real numbers, and $\tilde{\mathbf{U}}$ and $\tilde{\mathbf{V}}$ are $L_r \times L_r$ arbitrary unitary matrices. Note that $\lVert \mathbf{W}_{\text{BB},p} \rVert_F^2$ is not constrained as it does not change receive SNR of the system.

With the constraint $\tilde{\mathbf{\Sigma}}_p$ being a diagonal matrix with non-negative entries, the only optimal solution that makes the cost function is the scaled identity matrix, in other words, $\tilde{\mathbf{\Sigma}}_p = \sqrt{N_r/G} \mathbf{I}_{L_r}$ for all $p$. 
Then the digital combiner codebook can be defined as
\begin{align}
  \label{eq:Wbb_set}
  \mathcal{W}_{\text{BB}} &= \bigg\{ \mathbf{W}_{\text{BB},p} \in \mathbb{C}^{L_r \times L_r}: \forall p \in \{ 1, \ldots, L_r \}, \nonumber \\
  & \quad \mathbf{W}_{\text{BB},p} = \sqrt{{N_r}/{G}} \tilde{\mathbf{U}} \tilde{\mathbf{V}}^{\mathsf{H}} \bigg\} . 
\end{align}

\subsubsection{Minimizing $\nu(\tilde{\mathbf{W}})$}
\label{subsub:minimizing_nu_W}
As with Section~\ref{subsub:minimizing_nu_X}, $\min \nu( \tilde{\mathbf{W}} )$ is equivalent to $\min \sum_{m}^{G} \lvert \tilde{\mathbf{W}}(m)^{\mathsf{H}} \allowbreak \tilde{\mathbf{W}}(m) \rvert$, and its cost function can be expressed as $\text{Tr}(\tilde{\mathbf{W}} \tilde{\mathbf{W}}^{\mathsf{H}})$. Since the analog combiners have a condition $\mathbf{W}_{\text{RF},p}^{\mathsf{H}} \mathbf{W}_{\text{RF},p} = \mathbf{I}_{L_r}$ and the digital combiner codebook has a single element, $\text{Tr}(\tilde{\mathbf{W}} \tilde{\mathbf{W}}^{\mathsf{H}})$ reduces to $G/N_r \text{Tr}(\mathbf{W}_{p}^{\mathsf{H}} \mathbf{W}_{p})$ which is $G/N_r \text{Tr}(\tilde{\mathbf{\Sigma}}^{\mathsf{H}}_{p} \tilde{\mathbf{\Sigma}}_{p})$ assuming $\mathbf{W}_{\text{BB},p} = \tilde{\mathbf{U}} \tilde{\mathbf{\Sigma}}_{p} \tilde{\mathbf{V}}^{\mathsf{H}}$. A trivial solution to $\min \text{Tr}(\tilde{\mathbf{\Sigma}}^{\mathsf{H}}_{p} \tilde{\mathbf{\Sigma}}_{p})$ is $\tilde{\mathbf{\Sigma}}_p = \mathbf{0}_{L_r}$, which does not make sense. We thus pose a temporary constraint $\lVert \mathbf{W}_{\text{BB},p} \rVert_F^2 = \alpha > 0$. Then the problem now becomes
\begin{align}
  \min \text{Tr}(\tilde{\mathbf{\Sigma}}^{\mathsf{H}}_{p} \tilde{\mathbf{\Sigma}}_{p}) \text{ subject to } \lVert \tilde{\mathbf{\Sigma}}_{p} \rVert_F^2 = \alpha > 0. \nonumber
\end{align}
The optimal $\tilde{\mathbf{\Sigma}}_{p}$ can be found as $\sqrt{\alpha / L_r} \mathbf{I}_{L_r}$, which leads to $\mathbf{W}_{\text{BB},p} = \sqrt{\alpha / L_r} \tilde{\mathbf{U}} \tilde{\mathbf{V}}^{\mathsf{H}}$. Considering the fact that an arbitrary scaler is acceptable, the digital combiner that minimizes $\mu^t(\tilde{\mathbf{W}})$ also minimizes $\nu(\tilde{\mathbf{W}})$.

\section{Hybrid Architectures} 
\label{sec:hybrid_architectures}
In this section, we introduce some hybrid architectures and show applicability of the proposed sensing matrix design to them. Codebook conditions that we derived in the previous section with respect to the analog beamformers can be expressed in a single equation as
$\mathbf{F}_{\text{RF}}^{\mathsf{H}} \mathbf{F}_{\text{RF}} = \mathbf{I}_{N_t}$ where $\mathbf{F}_{\text{RF}} \triangleq [\mathbf{F}_{\text{RF},1}, \ldots, \mathbf{F}_{\text{RF}, N_t / L_t}]$. Any hybrid beamforming architecture whose analog precoders/combiners satisfy this condition can directly adopt the proposed deterministic configuration. 

\begin{figure}[t!]
  \centering
  \subfloat[]{
    \includegraphics[width=0.8in]{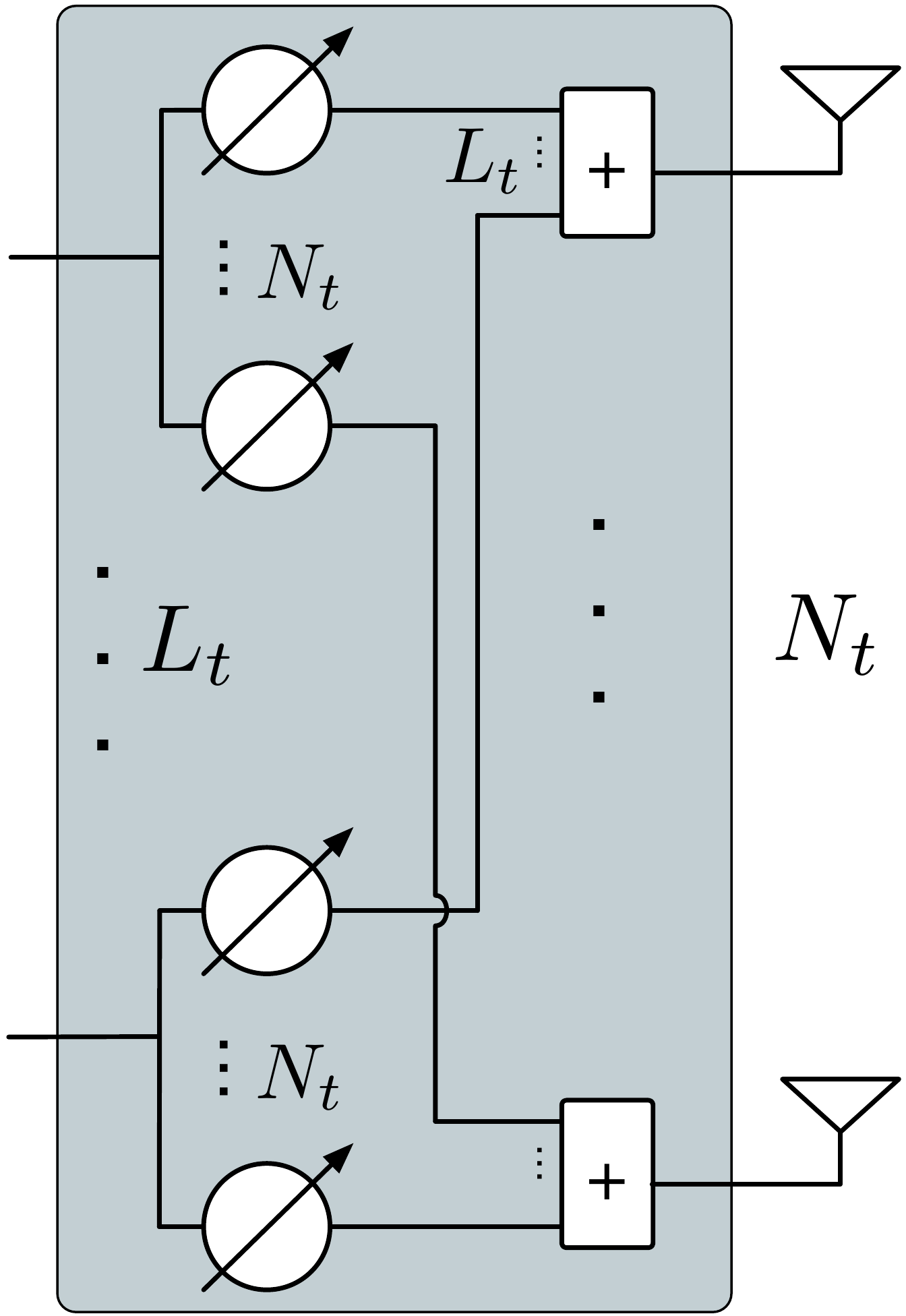}
    \label{fig:hybrid_arch_a}
  }
  \subfloat[]{
    \includegraphics[width=0.8in]{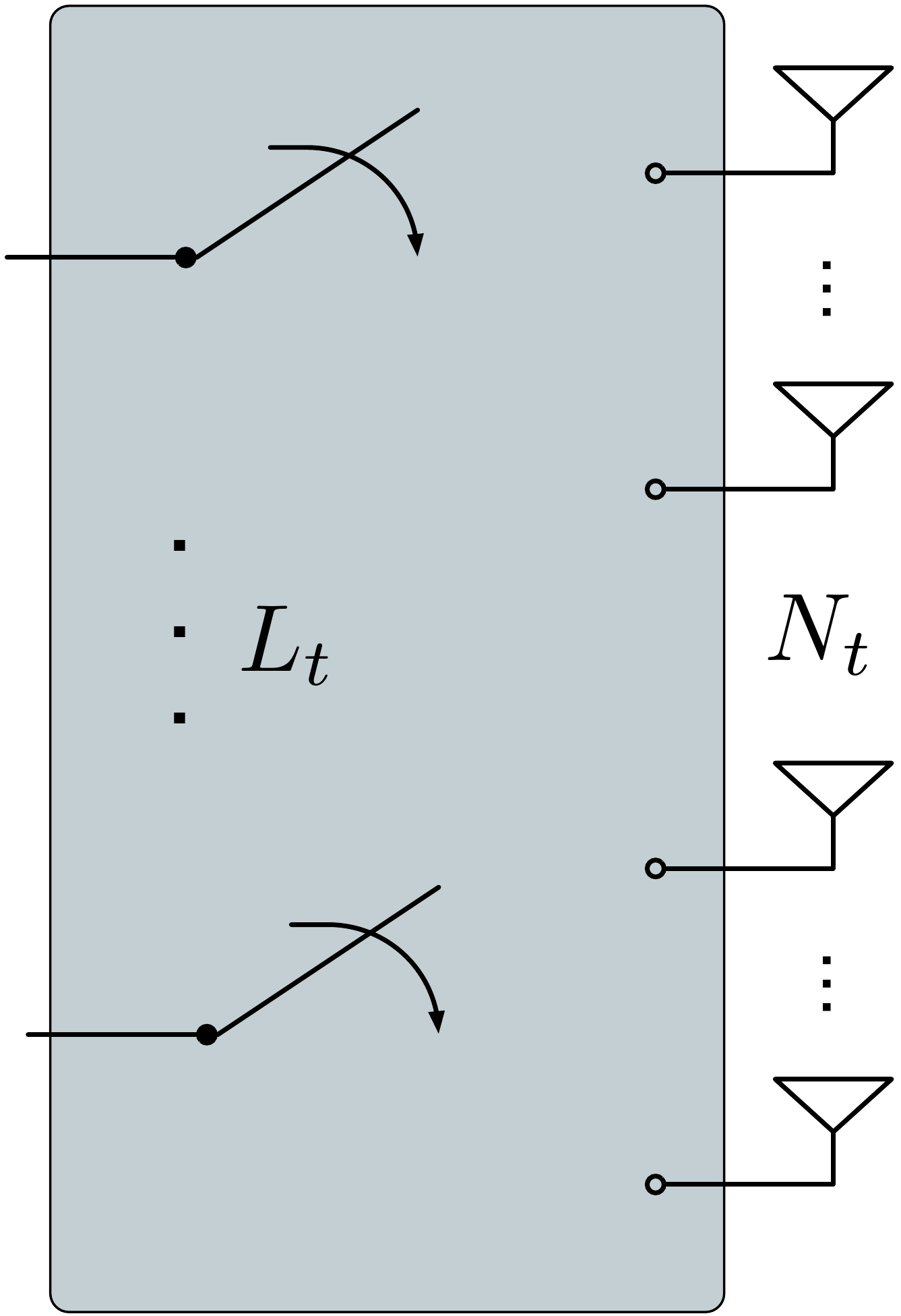}
    \label{fig:hybrid_arch_e}
  }
  \subfloat[]{
    \includegraphics[width=0.8in]{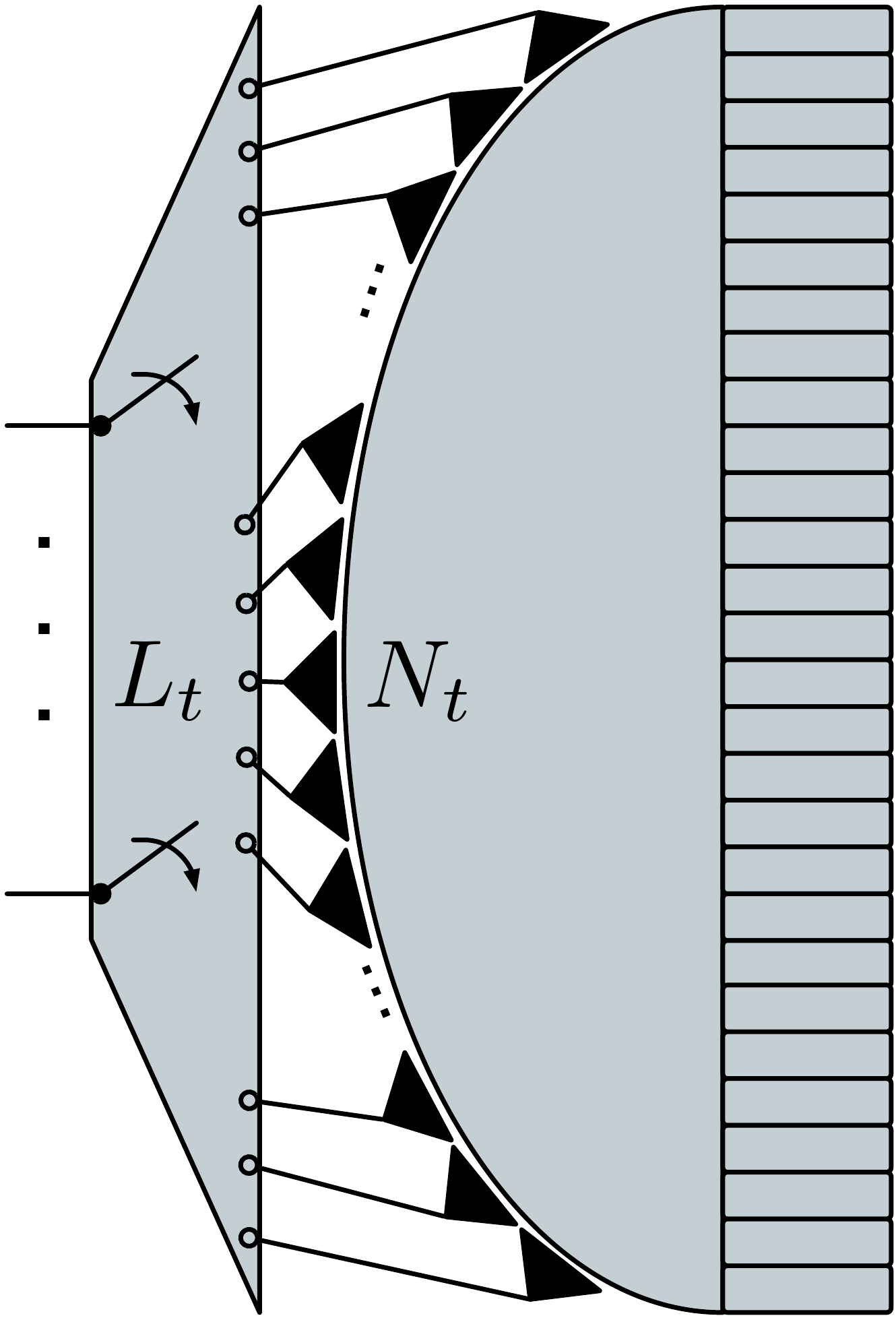}
    \label{fig:hybrid_arch_lens}
  }
  \caption{
  Three types of analog precoders: (a) phase shifting network, (b) switching network, and (c) CAP-MIMO.
  }
  \label{fig:hybrid_architectures}
  \vspace*{-0.1in}
\end{figure}

Promising architectures considered for hybrid beamforming are based on either networks of variable phase shifters, networks of switches or a DLA as illustrated in Fig.~\ref{fig:hybrid_architectures}
We define a set of feasible analog precoding vectors $\mathcal{F} \subset \mathbb{C}^{N_t \times 1}$ with different subscripts for different architectures. 

\subsection{Phase Shifting Network} 
\label{sub:phase_shifting_network}
In this architecture, each RF chain is connected to all antennas via a network of phase shifters as shown in Fig.~\subref*{fig:hybrid_arch_a}. Each network is composed of $N_t$ variable phase shifters, and there exists $L_t$ such networks. 
In total, $N_t L_t$ variable phase shifters are used.
Due to the hardware constraint of phase shifters, the set of feasible precoding vectors is given as
  $\mathcal{F}_1 = \left\{ \mathbf{f} \in \mathbb{C}^{N_t} : \lvert \mathbf{f}_i \rvert = \sqrt{1/N_t}, \angle \mathbf{f}_i \in \Theta \right\} \nonumber $
where $\Theta \triangleq \{ \theta: \theta = 2 \pi n / 2^{b_{\text{PS}}}, n=0,\ldots,2^{b_{\text{PS}}}-1 \}$ is the possible quantized angle set.
Columns of the normalized $N_t$-point DFT matrix can be shown to be in $\mathcal{F}_1$. 
In this case, the cardinality $\lvert \mathcal{F}_1 \rvert$ is $N_t$, and $\mathbf{F}_{\text{RF}}$ with columns being all $\mathbf{f}\text{'s} \in \mathcal{F}_1$ satisfies $\mathbf{F}_{\text{RF}}^{\mathsf{H}} \mathbf{F}_{\text{RF}} = \mathbf{I}_{N_t}$. 

\subsection{Switching Network} 
\label{sub:switching_network}
The switching network architecture 
connects each RF chain to one of the antennas via a switch as shown in Fig.~\subref*{fig:hybrid_arch_e}. At a given moment, the number of active antennas is $L_t$. The feasible set of precoding vectors is given as
  $\mathcal{F}_2 = \left\{ \mathbf{f} \in \mathcal{B}^{N_t}: \left\lVert \mathbf{f} \right\rVert_0 = 1 \right\} \nonumber $.
Due to the absence of splitters and combiners, the $\mathcal{L}_2$ norm of $\mathbf{f} \text{'s} \in \mathcal{F}_2$ is fixed at one. Without compensation from baseband processing, the condition is satisfied with a set of standard bases of $\mathbb{R}^{N_t}$ considered for the precoding vectors.  


\subsection{Continuous Aperture Phased MIMO} 
\label{sub:cap_mimo}
CAP-MIMO that directly exploits beamspace MIMO communications is enabled by high resolution DLAs \cite{brady2013tap}. For simulation, the DLA with adaptive selecting network proposed in \cite{gao2017twc} is used in the following section. A transmitter architecture of a 1D DLA is illustrated in Fig.~\subref*{fig:hybrid_arch_lens}. Feed antennas can ideally generate mutually orthogonal beams, and $L_t$ antennas out of $N_t$ are activated by selecting desired antennas and feeding input streams. 
To this end, DLAs are designed in order to make the analog precoding matrix approximate the DFT matrix. Considering the ideal precoding matrix, the set of analog precoding vectors can be expressed as
  $\mathcal{F}_{3} = \left\{ \mathbf{f} \in \mathbb{C}^{N_t}: \mathbf{f} = \mathbf{U}_{dft}(i), i=1, 2, \ldots N_t \right\} \nonumber$
where $\mathbf{U}_{dft}$ denotes the $N_t$-point DFT matrix. Since the DFT matrix is unitary, the analog precoding matrix obviously satisfies $\mathbf{F}_{\mathrm{RF}}^{\mathsf{H}} \mathbf{F}_{\mathrm{RF}} = \mathbf{I}_{N_t}$.

 

\begin{figure}[t!]
  \centering
  \centerline{
	  \subfloat[ Grid size $G$ is 64 points ]{
	    \includegraphics[width=3.1in]{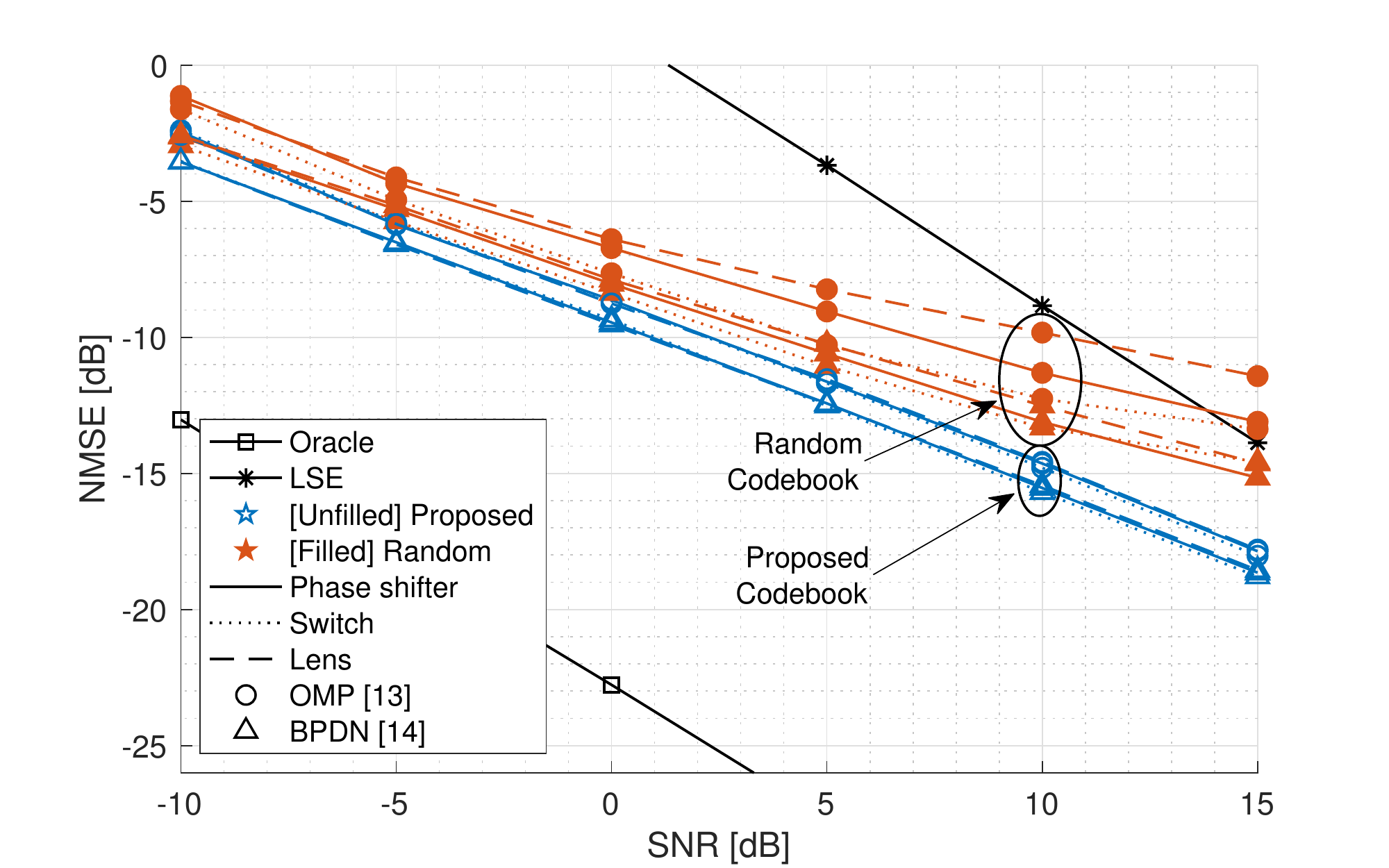}
	    \label{fig:nmse_Lr8_G64}
	  }  
  }
  \hfil
  \centerline{
		\subfloat[ Grid size $G$ is 180 points ]{
	    \includegraphics[width=3.1in]{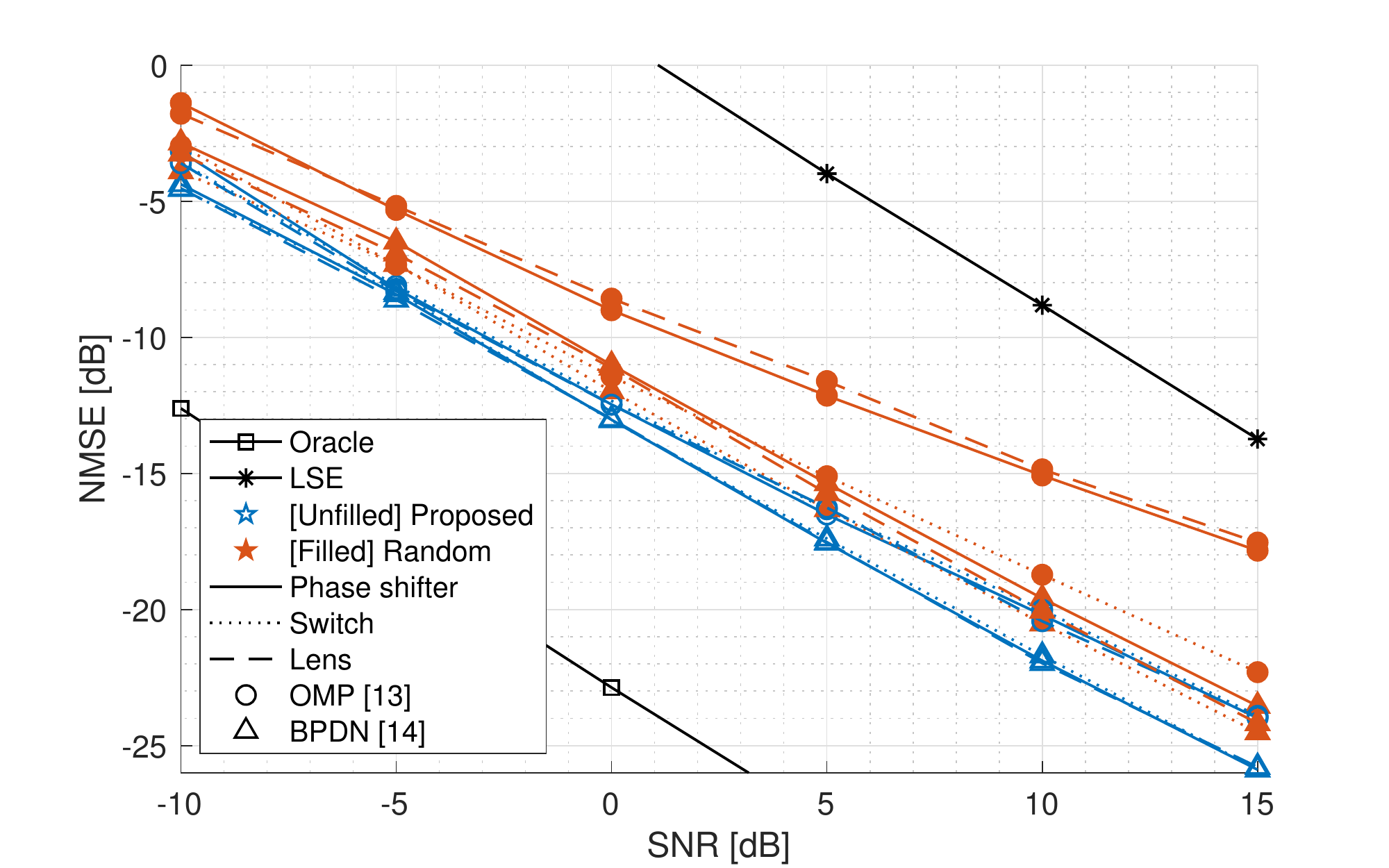}
	    \label{fig:nmse_Lr8_G180}
	  }  
  }
  \caption{
  NMSE for channel estimation vs. SNR.  The Oracle estimator knows the AoA and AoD.  The LSE uses the proposed deterministic sensing matrix.  The other results are for the possible combinations of deterministic/random hybrid beamforming designs, three RF beamforming architectures, and two estimation algorithms.
  }
  \label{fig:nmse_rf_chain}
  \vspace*{-0.2in}
\end{figure}

\begin{figure}[t!]
  \centering
  \centerline{
	  \subfloat[ Grid size $G$ is 64 points ]{
	    \includegraphics[width=3.1in]{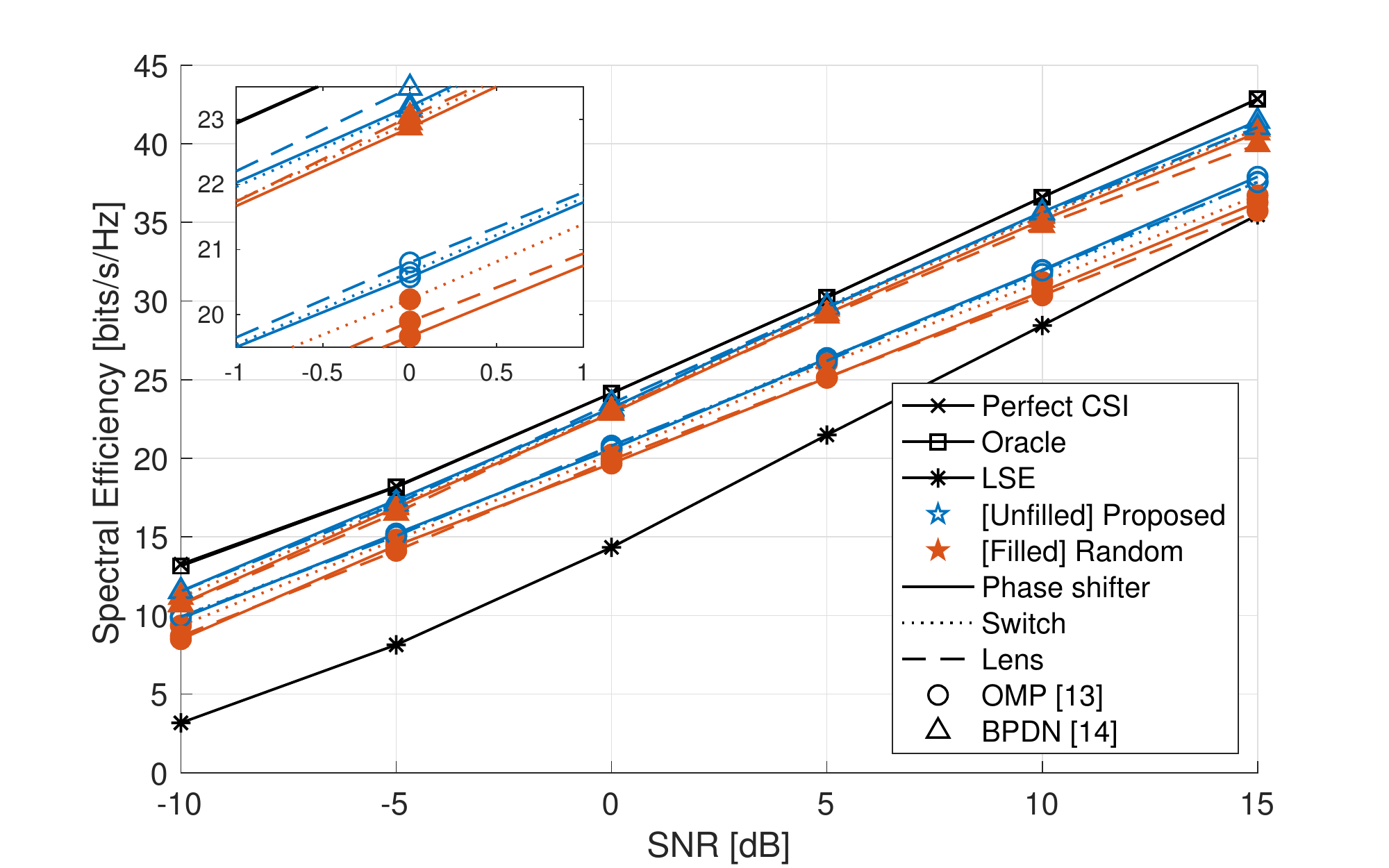}
	    \label{fig:SE_G64_Lr28}
	  }
  }
  \hfill
  \centerline{
	  \subfloat[ Grid size $G$ is 180 points ]{
	    \includegraphics[width=3.1in]{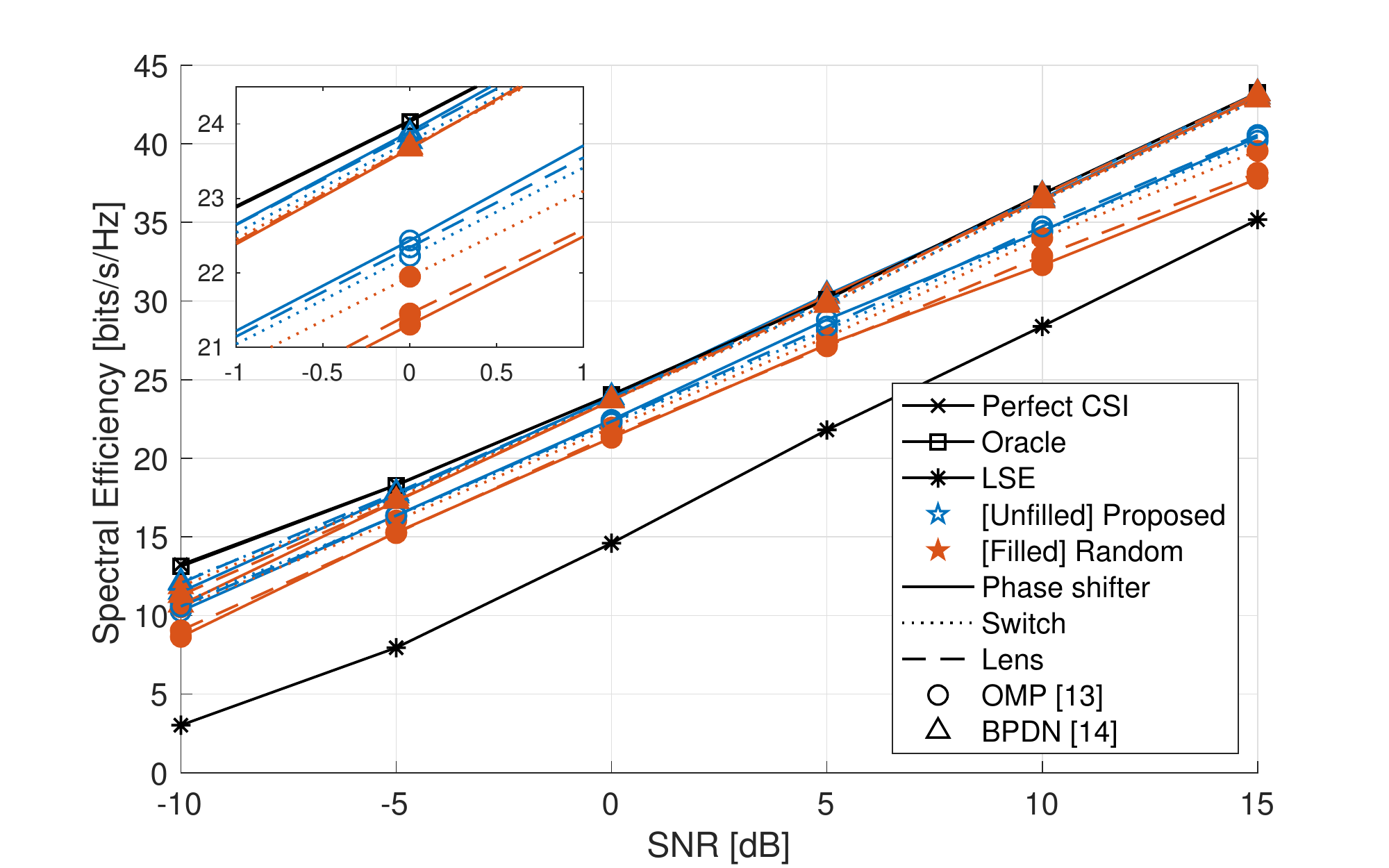}
	    \label{fig:SE_G180_Lr28}
	  }
  }
  \caption{
  SE for channel estimation vs. SNR. The Oracle estimator knows the AoA and AoD.  The LSE uses the proposed deterministic sensing matrix. For the possible combinations of hybrid beamforming designs, RF precoders, and estimation algorithms, the SE is computed using SVD beamforming.
  } 
  \label{fig:SE}
  \vspace*{-0.2in}
\end{figure}

\section{Numerical Results} 
\label{sec:numerical_results}
In this section, performance of CE based on the proposed sensing matrix is evaluated using OMP and BPDN algorithms as representatives of CS algorithms with coherence-based recovery guarantees \cite{duarte2011tsp}. Both the normalized mean squared error (NMSE) and the achievable SE are used as performance metrics where NMSE is defined as $\mathbb{E}[ \lVert \mathbf{H} - \hat{\mathbf{H}} \rVert_F^2 / \lVert \mathbf{H} \rVert_F^2]$. 
SNR is defined as $\rho / \sigma^2$.
We provide results obtained with the random sensing matrix in addition to the one we propose. For comparison purposes, the least squares estimator (LSE) and the oracle estimator are evaluated as well. The LSE evaluated in this section is based on the proposed sensing matrix. The oracle estimator refers to the LSE with actual AoAs and AoDs known at the receiver. 

The system is equipped with 
$N_t=64$, $N_r=16$, $L_t=8$, $L_r=8$ (hence $M_t=64$ and $M_r=2$), $G=64$, $N_p=4$ and $b_{\text{PS}}=6$
for simulation unless otherwise specified. 
The AoDs and AoAs of the multipath components are not constrained to lie on the angle grids of the dictionary. In the simulations, 500 channel realizations are used for each point. The complete source code is available \cite{projectcode}.
Fig.~\ref{fig:nmse_rf_chain} shows NMSE of channel estimates as a function of SNR with combinations of codebooks, architectures and algorithms.
The oracle estimator outperforms the others and scales well with SNR serving as the lower bound.
Both figures show that the proposed deterministic design outperforms the random design for all considered combinations across the entire SNR range including the low SNR regime. Performance of the OMP and BPDN with the random sensing matrix, in general, is between the proposed design and the LSE.  
It is expected because the equivalent dictionary has higher total coherence than the proposed one does. 
Compared with the random design, the proposed deterministic design makes the various HB architectures achieve very similar performance since they generate dictionaries with the identical total coherence as shown in Section~\ref{sec:hybrid_architectures}. 
Spectral efficiencies computed based on SVD beamforming for various combinations of architectures and estimation algorithms are plotted in Fig.~\ref{fig:SE}. The SE using the perfect channel state information (CSI) is provided in the figure as the performance upper bound. In both figures, the oracle estimator yields the indistinguishable SE from the perfect CSI since NMSE is very low across the SNR range as shown in Fig.~\ref{fig:nmse_rf_chain}. The proposed deterministic design outperforms the random from the perspective of SE as well, and makes spectral efficiencies of the various architectures tend to converge.

In both Fig.~\ref{fig:nmse_rf_chain} and \ref{fig:SE}, an increase in the grid size is favorable for both deterministic and random approaches since larger grids mean finer resolution in the angle search spaces. By increasing $G$ from $64$ to $180$, NMSE curves are shifted down and SE curves are lifted up for both OMP and BPDN. At the same time, however, gaps between the proposed and the random become smaller. It implies that the proposed codebook is more efficient when with a smaller grid size which could be preferred for computation reduction.

\section{Conclusion} 
\label{sec:conclusion}
In this paper, we proposed a versatile deterministic HB design framework for CS based CE in narrowband mmWave communication systems. The deterministic sensing matrix design we proposed works for a variety of hybrid beamforming architectures that are implemented with variable phase shifters, switches or a DLA. Our design approach is to configure analog and digital beamformers by minimizing the total coherence of the equivalent sparsifying dictionary. 
We decoupled the joint transmitter and receiver optimization problem into two disjoint problems. 
The analog and digital beamformer codebooks that are obtained by solving the optimization problems improve CE performance of the CS algorithms that rely on coherence to guarantee sparse recovery and were shown by simulation to outperform, in terms of both estimation error and SE. 
\bibliographystyle{IEEEtran}
\bibliography{ieeeabrv,paper}
\end{document}